\documentclass[a4paper,12pt]{article}
\usepackage{graphicx}
\usepackage{subfigure}
\usepackage{caption}
\usepackage{color}

\DeclareGraphicsExtensions{.epsi,.eps,.ps}

\begin{document}
\title{Quantum Communication in Spin Systems With Long-Range Interactions}
\author{M. Avellino, A.J. Fisher, S. Bose\\Department of Physics and Astronomy, and\\London Centre for Nanotechnology\\University College London, Gower Street, London WC1E
6BT}
\date{14 March 2006}
\maketitle
\begin{abstract}
We calculate the fidelity of transmission of a single qubit between
distant sites on semi-infinite and finite chains of spins coupled
via the magnetic dipole interaction. We show that such systems often
perform better than their Heisenberg nearest-neighbour coupled
counterparts, and that fidelities closely approaching unity can be
attained between the ends of finite chains without any special
engineering of the system, although state transfer becomes slow in
long chains. We discuss possible optimization methods, and find
that, for any length, the best compromise between the quality and
the speed of the communication is obtained in a nearly uniform chain
of 4 spins.
\end{abstract}
\section{Introduction}
Quantum Information Processing (QIP) offers several advantages over
classical computation in solving problems associated with large
and/or dynamical systems. One of the main goals of current research
in the field is to find ways and means of reliably transmitting
quantum information, encoded in quantum bits (or \emph{qubits}),
over arbitrarily large distances. In order to ensure the information
is received as the sender intended, the qubits must be protected
from interacting with the environment in any way. For this reason,
until now the qubits of choice have mostly been photons, which have
an extremely small interaction cross section and can thus be used to
generate quantum states that are sufficiently robust to perform
protocols such as quantum cryptography and teleportation
(\cite{lim06,gisin02,bennett93,boschi98,furusawa98,bouwmeester97}).

However, in recent years much effort has been dedicated to studying
systems in which quantum information is encoded in \emph{stationary}
qubits, and is propagated from one part of the system to another by
the interaction between the system's components. One of the simplest
geometries in which this can be achieved is a one-dimensional chain
of interacting particles, where the qubit is encoded in some
internal degree of freedom (which we call `spin', using $|0 \rangle$
for spin `up' and $|1 \rangle$ for spin `down'). Originally an
exchange-coupled chain of spins with a constant nearest-neighbour
(nn) interaction was studied \cite{bose03}; it was found that a
qubit could be transferred with a fidelity exceeding the maximum
classical value in a time that grows polynomially with the length of
the chain. Subsequently, it was shown that such simple chains allow
transmission fidelities arbitrarily close to unity also if the qubit
is taken to be a carefully designed `wave packet', provided the
sending and receiving parties can access a sufficiently large
portion of the chain \cite{osborne04}. Stronger results can be found
for more complex systems: in the absence of structural
imperfections, an XY Hamiltonian on a hyper-cubic lattice allows
perfect state transfer \cite{christandl04,christandl05,dechiara05},
as does a pair of parallel spin chains \cite{burgarth05}, or a spin
chain acting as a quantum wire connecting two qubits
\cite{wojcik05}. Plenio \textit{et al.} \cite{plenio05} have studied
the situation for chains of harmonic oscillators (i.e. where each
particle on the lattice possesses a continuous, rather than a
discrete, degree of freedom), while Hartmann \textit{et al.}
\cite{hartmann05} have recently found that quantum information can
be made to propagate with arbitrarily high fidelity through both
oscillator and spin chains near a quantum phase transition, provided
the ground state and the lowest excited state of the system are not
degenerate. However, this transfer is exponentially slow; more rapid
transmission is possible at the quantum critical point, but at some
cost to the fidelity.

It has also been shown that high fidelities can be attained by
engineering the strength and the nature of the interactions between
the spins \cite{yung05}. However, this requires structures that
would be very difficult to manufacture, both because the
communicating parties would need to have an extremely high degree of
control over the system, and because the component spins would have
to exhibit nn couplings only, with precisely defined strengths. This
is clearly an idealization, because long-range interactions are also
likely to be present. Previous work has been dedicated to systems
more `realistic' from this point of view. Kay \textit{et al.} have
studied finite spin chains in which the total Hamiltonian accounts
for the presence of local magnetic fields and a potential of the
form of the magnetic dipole interaction \cite{kay05}. The approach
adopted in this case was to pre-determine a spectrum of eigenvalues
that would ensure perfect state transfer for all chain lengths, and
subsequently derive the corresponding local fields and the
inter-spin distances by solving an inverse eigenvalue problem. This
method is applicable to a Hamiltonian containing any number of
parameters, and could provide very useful theoretical guidelines on
the optimal way to structure a system, although, once again, the gap
between theory and experiment may prove difficult to bridge.

In this work, by contrast, we investigate simple one-dimensional
arrays of spins interacting via a pure magnetic dipole interaction.
We allow no site-specific locally-tunable fields; nevertheless, we
show that fidelities for quantum state transfer closely approaching
unity can be attained between the ends of finite chains, without any
special engineering of the system. Furthermore, because of the long
range interaction, the transfer rate grows polynomially in the
system size, rather than exponentially, as in the case studied by
Hartmann \textit{et al.} \cite{hartmann05}. Our results may be
relevant to two-level atoms in atomic traps \cite{micheli05,polar04}
or to one-dimensional arrays of endohedral fullerene species
encapsulated within carbon nanotubes \cite{khlobystov04}, as well as
to natural magnetic dipolar systems such as $\mathrm{LiHoF_4}$
\cite{bitko96}, and finite spin chains of a more complex design, for
example engineered from arrays of quantum dots \cite{damico05}.

\section{The System}
\label{The System} We build on the work done by Bose \cite{bose03}
on transferring quantum information through an infinite, uniform
chain exhibiting isotropic nn interactions only. In this system, the
qubit is represented by a single flipped spin, which propagates
between different sites in a manner defined by a time-independent
Hamiltonian of the form:
\begin{equation}
H = -\frac{J}{2}\delta_{i + 1,j}\sum_{\langle i,j \rangle}
\mathbf{\sigma}^i \cdot \mathbf{\sigma}^j -B\sum_{i =
1}^N\mathbf{\sigma}^i_z
\end{equation}
where \textit{N} is the number of spins in the chain,
$\mathbf{\sigma^i} = (\sigma^i_x,\sigma^i_y,\sigma^i_z,)$ are the
Pauli spin matrices for the $i^{th}$ spin, $B > 0$ is a uniform
magnetic field, and $(J/2)\delta_{i + 1,j}$ is the coupling strength
between spins \textit{i} and \textit{j}, which is non-zero for
nearest-neighbouring spins only. If the ground state of the system
is expressed as $|\downarrow \rangle \bigotimes|\downarrow \rangle
\bigotimes...\bigotimes |\downarrow \rangle$, it has been shown
that, provided the chain is sufficiently short, perfect or near
perfect state transfer can be achieved by simply letting an initial
state of the form $|\uparrow \rangle\bigotimes|\downarrow
\rangle\bigotimes...\bigotimes |\downarrow \rangle$ evolve naturally
in time according to the effects of \textbf{H}. It is important to
note that $[H ,\sum_{i = 0}^N \sigma_z^i] = 0$, that is, the
Hamiltonian conserves the total magnetization \textbf{M} of the
system, allowing the chosen initial state to evolve only into states
in which \emph{one} spin is flipped at any given time.

We propose to investigate the quality and efficiency of quantum
state transfer through infinite and finite chains of
spin-$\frac{1}{2}$ fermions coupled by long-range interactions
having the form of the magnetic dipole interaction. These two
systems differ in that the finite chain has end points, whereas the
infinite chain does not; indeed, due to the periodic nature of the
infinite chain we will hereafter refer to it as a \emph{ring}. We
examine a simplified system in which any external magnetic field is
constant and parallel to the axis joining the dipoles, which is
chosen to coincide with the \textit{z} direction. The alignment of
the magnetic and dipole axes is not a trivial point, but a necessary
condition to ensure that the total magnetization \textbf{M} remains
a good quantum number, and allows us to work in the sub-space where
only one spin is flipped with respect to the ground state, reducing
the Hamiltonian from a $2^N \times 2^N$ to a $N \times N$ matrix.
Within this subspace the effect of the magnetic field is to add a
constant to the energies. We will hereafter omit this constant.
Following the notation used in \cite{bose03}, we denote with
$|0000...0 \rangle$ the (unique) ground state of the system (i.e.
all spins facing down, parallel to the external field) and with $|j
\rangle$ the block of states in which the spin at the $j^{th}$ site
has been flipped from 0 to 1. For simplicity, we  assume there are
no thermally excited spin-flips in the system. We adopt a
Hamiltonian of the form:
\begin{equation}
H_d = \frac{C}{r^3}[\mathbf{S_i} \cdot \mathbf{S_j} - 3
\mathbf{S_i^z} \mathbf{S_j^z}]
\end{equation}
where \textit{C} is a constant, $\mathbf{S_i}$ and $\mathbf{S_j}$
are the total spin operators at sites \textit{i} and \textit{j}, and
$\mathbf{S_i^z}$ and $\mathbf{S_j^z}$ are the respective \textit{z}
components. The value of \textit{C} is determined by the type of
particle in the chain. For a system of spin-$\frac{1}{2}$ fermions
(e.g. electrons) \textit{C} is given by:
\begin{equation}
C = \frac{\mu_0(\mu_B g)^2}{4\pi \hbar^2}
\end{equation}
where $\mu_0$ is the permeability of free space, $\mu_B$ is the Bohr
magneton, \textit{g} is the electronic Land\`{e} g-factor and $\hbar
= h/2 \pi$. Throughout this paper we will assume that $\mu_0 = \mu_B
= \hbar = 1$, so that:
\begin{equation}
C = \frac{g^2}{4\pi}
\end{equation}
We define \textit{a} to be the spacing between neighbouring
fermions. In this case, the strength of the interaction between
nearest neighbours is:
\begin{equation}\label{eqnnint}
\langle i|H_d|i \pm 1 \rangle = \frac{C}{2a^3}
\end{equation}
For the results shown we define our length, energy and time units by
setting the nearest-neighbour separation and the interaction energy
between nearest neighbours to unity. However, eqn.(\ref{eqnnint})
implies the Hamiltonian has an overall scaling factor of $1/a^3$, so
a uniform compression or expansion of the system should only have
the \emph{quantitative} effect of re-scaling the system's energy by
a constant. Therefore, provided the chain remains uniform and the
number of component spins is fixed, the energy and performance of a
chain of any size can be extrapolated by simply adjusting the value
of \textit{a} as necessary.

\subsection{Rings}
We initially consider a ring of \textit{N} spins in its ground
state. Our aim is to calculate the maximum fidelity of transmission
of a qubit from site \textit{r} to a distant site \textit{s}, as a
function of time and number of spins in the ring. We denote the
initial and final states of the system by $|r \rangle$ and $|s
\rangle$ respectively. The expression for the maximum fidelity of
quantum state transfer is given by \cite{bose03} as:
\begin{equation}
F_{r,s}^N(t) = \frac{ \left| f_{r,s}^N \left( t \right) \right|}{3}
+ \frac{ \left| f_{r,s}^N \left(t \right) \right| ^2}{6} +
\frac{1}{2}
\end{equation}
where $f_{r,s}^N \left( t \right)$ is the propagator, which is
calculated from the following:
\begin{equation}
f_{r,s}^N \left( t \right) = \sum_{m = 1}^N \langle r|m \rangle
\langle m|s \rangle e^{-i E_m t}
\end{equation}
We assume that the eigenvectors $|m \rangle$ of the system can be
expanded using the basis formed by the $|j \rangle$ states. Imposing
periodic boundary conditions allows us to express these eigenstates
as Bloch states, so that for a ring of circumference $L = Na$:
\begin{equation}\label{eqkexpansion}
|m \rangle = \frac{1}{\sqrt{N}}\sum_{j = 0}^{N - 1} e^{ik_mja}|j
\rangle
\end{equation}
The propagator can then be re-expressed as:
\begin{equation}
f_{r,s}^N (t) = \frac{1}{N} \sum_{m=1}^Ne^{ik_ma \left( r-s \right)}
e^{- i E_m t}=\frac{1}{N} \sum_{m=0}^{N-1}e^{i\frac{2 \pi m}{N}
\left( r-s \right)} e^{- i E_m t}
\end{equation}
Using eqn. (\ref{eqkexpansion}), we calculate the energies of the
system, which can be written as:
\begin{equation}
E_m = \langle m|H_d|m \rangle = \frac{1}{N}\sum_{i,j}^{N - 1}
e^{ik_ma \left( j - i \right)} \langle i |H_d| j \rangle
\end{equation}
We require only matrix elements for which $i \not= j$, since by
symmetry the diagonal terms of $H_d$ are independent of \textit{N},
and therefore change the energy of the system by a constant shift.
For evenly spaced spins with nearest-neighbour separation
\textit{a}, these off-diagonal terms are:
\begin{equation}\label{eqoffdiagring}
\langle i|H_d|j \rangle = \frac{C}{2|r_j - r_i|^3} = \frac{C}{2a^3
|j - i|^3}
\end{equation}
As eqn. (\ref{eqoffdiagring}) only depends on the difference $|j -
i|$, we can reduce the double summation to a single sum over
\textit{j} by fixing a value of \textit{i}. For convenience we
choose $i = 0$, so that:
\begin{equation}
E_m = \frac{C}{2a^3} \sum_{j = -
\frac{N}{2}}^{\frac{N}{2}}e^{ik_mja} \frac{1}{|j^3|} = \frac{C}{a^3}
\sum_{j = 1}^{\frac{N}{2}}\frac{\cos{\left(k_mja\right)}}{j^3}.
\end{equation}
\begin{figure}[h]
\renewcommand{\captionfont}{\footnotesize}
\renewcommand{\captionlabelfont}{}
\begin{center}
\subfigure{\label{mfr}\includegraphics[width=6cm]{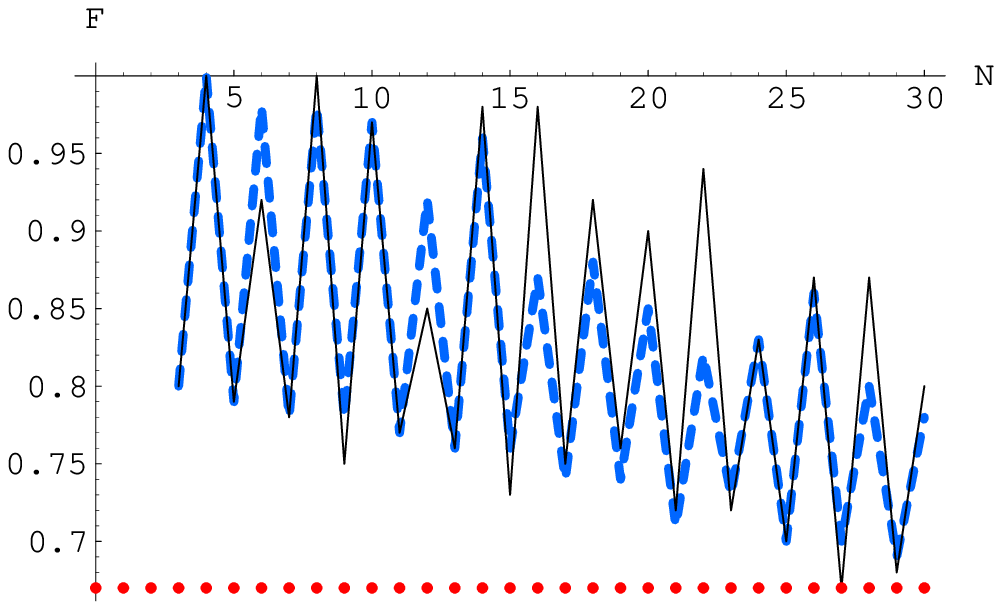}}
\hspace{0.3cm}
\subfigure{\label{timerings}\includegraphics[width=6cm]{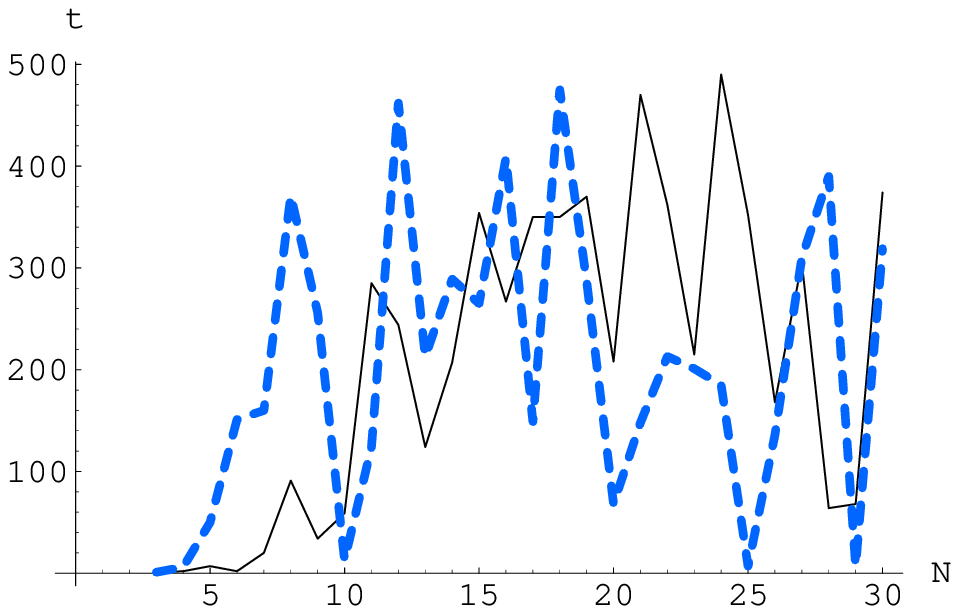}}
\end{center}
\caption{(Colour online) Figure (a) shows the maximum fidelity that
can be achieved in transferring an input state $|1 \rangle$ to an
output state $|\frac{N}{2}+1\rangle$ or $|\frac{N + 1}{2}\rangle$ on a
ring of spins coupled by dipole-dipole (dashed blue curve) or
nearest-neighbour (black curve) interactions, as a function of
\textit{N}. The red dotted line at $F = 2/3$ indicates the highest
fidelity for classical transmission of a quantum state. Figure (b)
shows the time at which the fidelity first peaks in this system. Units are as specified in Section
\ref{The System}.} \label{rings}
\end{figure}

Fig.~\ref{mfr} shows the maximum fidelity of state transfer for
rings of $N = 3$ to $N = 30$ spins, when the sending and receiving
parties are located at diametrically opposite sites, for which $|r
\rangle = |1\rangle$ and $|s \rangle = |\frac{N}{2}+1\rangle$ or $|s\rangle = |\frac{N
+ 1}{2}\rangle$ for even and odd \textit{N} respectively. We have assumed
that transfer occurs along the arc joining sites \textit{r} and
\textit{s}, though this may not be exactly the case. Fig.~\ref{mfr}
also shows the performance of a ring in which the spins are coupled
by nn interactions only. We note that for $N > 3$ the performance of
the nn-coupled ring is slightly better unless $N=6$ or $N=12$. We also find that the times of optimum transfer
tend to rise as we increase \textit{N} in both the dipole and the
nn-coupled rings.

\subsection{Single Qubit Transfer in Uniform Chains}
We now extend the previous analysis to a finite chain, calculating
the full Hamiltonian of the system, which has the form:
\begin{equation}\label{eqoffdiag}
\langle i|H_d|j \rangle = \frac{C}{2a^3|j - i|^3}
\end{equation}
\begin{equation}\label{eqdiag}
\langle j|H_d|j \rangle = -\frac{C}{2a^3}\sum_{\langle k,l
\rangle}\frac{1}{{|k - l|}^3} + \frac{C}{a^3}\sum_{i \not=
j}\frac{1}{{|j - i|}^3}
\end{equation}
where $\langle 000..0|H_d|000..0 \rangle =
-\frac{C}{2a^3}\sum_{\langle k,l \rangle}\frac{1}{{|k - l|}^3}$ is
the ground state energy of the system.

\begin{figure}[h]
\renewcommand{\captionfont}{\footnotesize}
\renewcommand{\captionlabelfont}{}
\begin{center}
\includegraphics[width=7cm]{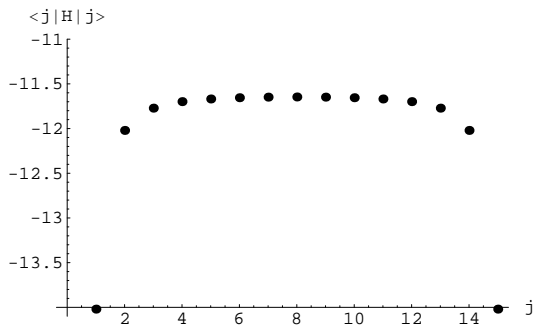}
\end{center}
\caption{The on-site energy as a function of the site \textit{j} of
the spin flip for a chain of 15 magnetic dipole-coupled spins. We
note that the energies at sites 1 and 15 are much lower than the
rest. Units are as specified in Section \ref{The System}.}
\label{diagelnine}
\end{figure}

\begin{figure}[h]
\renewcommand{\captionfont}{\footnotesize}
\renewcommand{\captionlabelfont}{}
\begin{center}
\includegraphics[width=7cm]{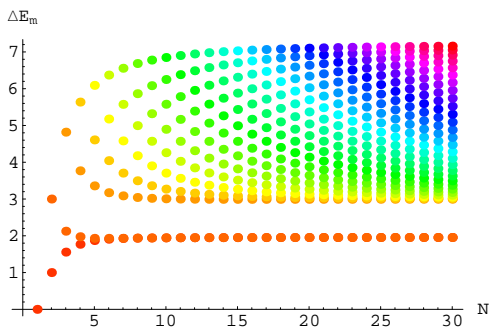}
\end{center}
\caption{(Colour online) The energy splitting $\Delta E_m$ between
the ground state $|0000..0\rangle$ and states with a single flipped
spin for a uniform chain of \textit{N} spins, showing the evolution
of the two bound states. Values of $\Delta E_m$ of the same index
\textit{m}, counting from the bottom of the spectrum, are shown in
the same colour. Units are as specified in Section \ref{The
System}.} \label{spectrum}
\end{figure}

The fidelity of state transfer between the ends of the chain is
obtained by taking $|r \rangle = |1 \rangle$ and $|s \rangle = |N
\rangle$. We find that $F_{1,N}^{N}(t)$ exhibits three `trademark'
features.
\begin{figure}[h]
\renewcommand{\captionfont}{\footnotesize}
\renewcommand{\captionlabelfont}{}
\begin{center}
\includegraphics[width=7cm]{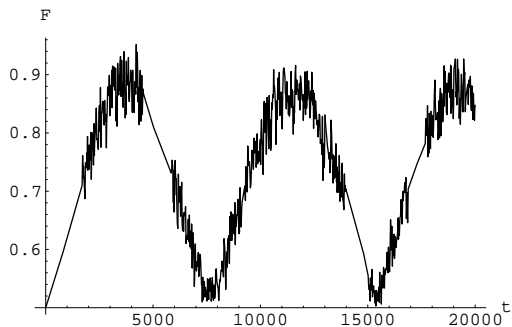}
\end{center}
\caption{The evolution in time (abscissa) of the fidelity of state
transfer between sites 1 and 10 of a uniform chain of magnetic
dipole-coupled spins. We note the regularity of the oscillation and
the high value of $F(max)$. Units are as specified in Section
\ref{The System}.} \label{nine}
\end{figure}\\First of all, the maximum value of $F(t)_{1,N}^{N}$ is close to unity. Secondly, the value of $F_{1,N}^{N}(t)$ oscillates
between 1/2 and the maximum (which we call $F_{max}$) with a regular
frequency, which is generally quite small, implying that state
transfer occurs slowly (fig.~\ref{nine}). Finally, the period of
oscillation of $F_{1,N}^{N}(t)$, which we call \textit{T}, is
uniquely defined by the energy splitting $\Delta \lambda$ between
the two lowest eigenvalues of $H_d(N)$. The transfer process is
therefore dominated by the beating of two nearly degenerate states
localized near the ends of the chain. This behaviour is explained by
the variation of the on-site energies of the spins as a function of
\textit{j}, shown in fig.~\ref{diagelnine}; it is immediately
evident that the most favourable positions for a spin to flip are
sites 1 and \textit{N}. Consequently, states $|1 \rangle$ and $|N
\rangle$ are the most strongly coupled to the system's (two) bound
states, which are shown in fig.~\ref{spectrum}. In this system, this
phenomenon is a natural consequence of the geometry, but systems in
which the spin flip energy is specifically chosen site by site have
also been studied \cite{santos05}.
\\The period of $F_{1,N}^N(t)$ is related to $\Delta \lambda$ by:
\begin{equation}
T = \frac{2 \pi}{\Delta \lambda}
\end{equation}
Consequently, the time at which $F_{1,N}^{N}(t)$ first peaks is:
\begin{equation}
t(F_{max}) = \frac{T}{2} = \frac{\pi}{\Delta \lambda}
\end{equation}
This time rises with chain length, as the splitting $\Delta \lambda$
decreases with increasing \textit{N}.

\begin{figure}[h]
\renewcommand{\captionfont}{\footnotesize}
\renewcommand{\captionlabelfont}{}
\begin{center}
\subfigure{\label{mfch}\includegraphics[width=6cm]{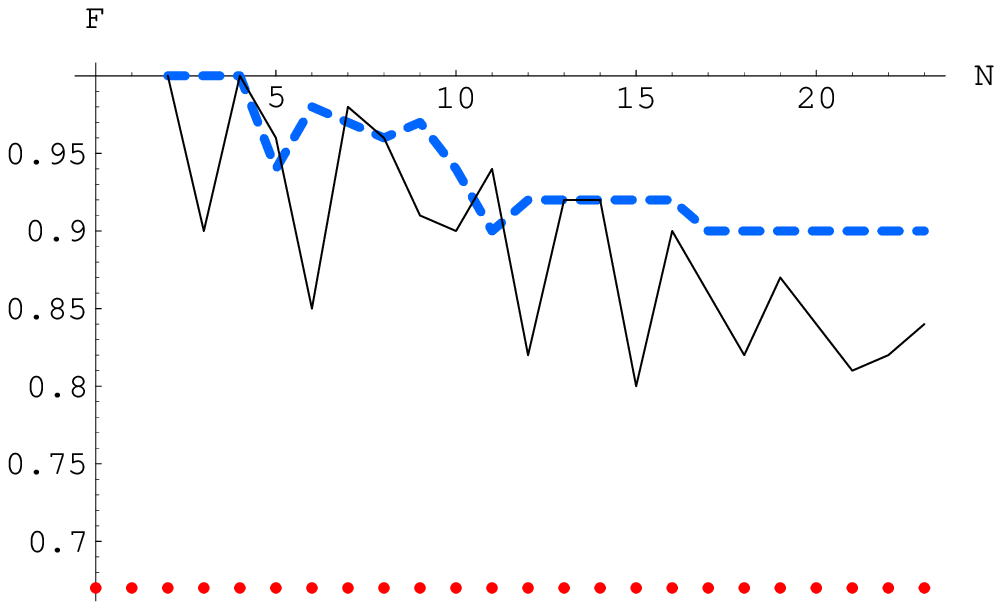}}
\hspace{0.3cm}
\subfigure{\label{timech}\includegraphics[width=6cm]{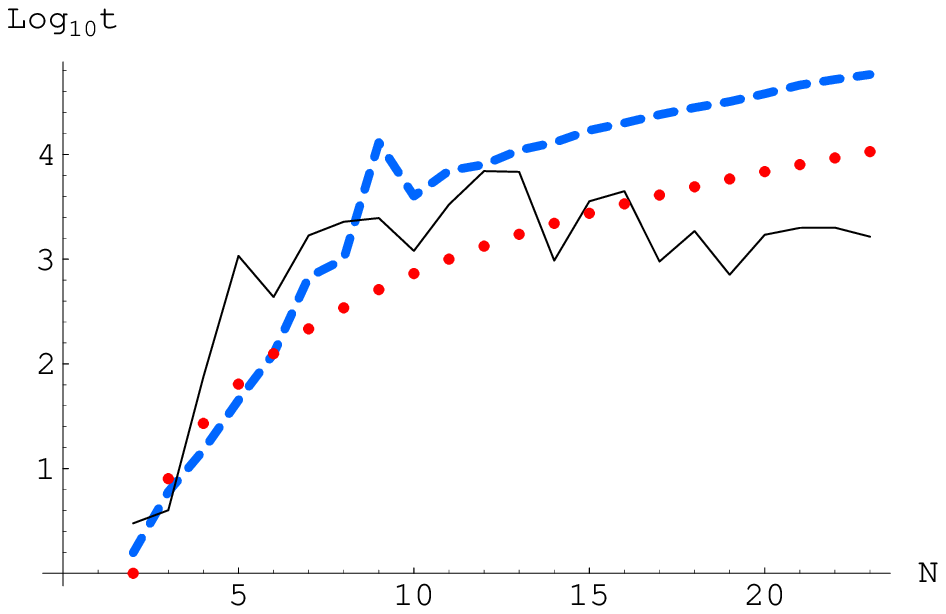}}
\end{center}
\caption{(Colour online) Figure (a) shows the maximum fidelity that
can be achieved in transferring an input state $|1 \rangle$ to an
output state $|N\rangle$ in a chain of spins coupled by
dipole-dipole (dashed blue curve) or nearest-neighbour (black curve)
interactions, as a function of \textit{N}. The red dotted line at $F
= 2/3$ indicates the highest fidelity for classical transmission of
a quantum state. We note that the dipole-coupled chain almost always
performs better. Figure (b) shows the time at which the fidelity
first peaks in these two systems, plotted on a $Log_{10}$ scale. The
red dotted curve is the function $y = L^3$. We see that at large
\textit{L} the dashed blue and red dotted curves are parallel,
indicating that the transfer time scales as the cube of the chain
length. Units are as specified in Section \ref{The System}.}
\label{chains}
\end{figure}

A summary of our results for $N = 2$ to $N = 23$ spins is shown in
Fig.~\ref{mfch}. We note that, in addition to $N = 2$, $N = 3$ and
$N = 4$ also give perfect transfer, and in general $F_{max} \geq
0.9$. This is a marked improvement on the performance of a
nn-coupled chain (also shown in Fig.~\ref{mfch}); in particular, it
seems that by replacing the nn couplings with dipole couplings we no
longer obtain poor transfer when \textit{N} is a multiple of 3
\cite{bose03}. Unfortunately, we again observe a `trade-off' between
fidelity and time, which becomes particularly evident for $N > 6$.
In fact, we find that at large \textit{N}, $t(F_{max})$ goes as the
cube of the chain length (fig.~\ref{timech}). It is therefore
evident that in long chains it will take an impractical length of
time to complete the protocol unless the system can be optimized in
some way.

\subsubsection{Structural Optimization} 
\begin{figure}[h]
\renewcommand{\captionfont}{\footnotesize}
\renewcommand{\captionlabelfont}{}
\begin{center}
 \includegraphics[width=7cm]{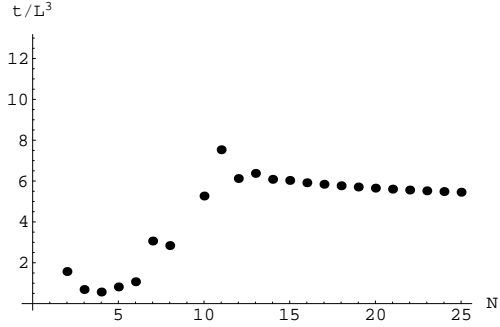}
 \end{center}
 \caption{The behaviour of $\tau$ as a function of the number
 of spins in the chain. Note the minimum at $N = 4$ and the flatness of
 the curve for $N > 15$. Units are as specified in Section
\ref{The System}.} \label{normtimech}
\end{figure}

We now analyze the
efficiency of state transfer for a \emph{fixed} chain length, as a
function of the number of spins in the chain. We define $\tau$ as
the transmission time giving maximum fidelity at unit chain length,
i.e.:
\begin{equation}
\tau = \frac{t(F_{max})}{L^3}
\end{equation}Fig.~\ref{normtimech} shows a plot of $\tau$ as a function of
\textit{N}, which reveals two interesting features. The first is the
presence of a minimum at $N = 4$, which we will discuss
subsequently. The second is the fact that $\tau$ tends to a constant
at large \textit{L}. This indicates that, above a certain threshold
value of \textit{N}, the evolution of the system is determined
almost exclusively by the magnetic dipole coupling between spins 1
to \textit{q} with spins $N - q$ to \textit{N}, irrespective of the
number of spins that separate these two `clumps'.
To explore this hypothesis, and determine the behaviour of $\tau$
for large \textit{N}, we work with states $|B \rangle$ and $|E
\rangle$ localized at the beginning and end of the chain, which are
the bound-state eigenfunctions of a semi-infinite chain extending to
the right and the left, respectively. We take:
\begin{equation}\label{eqb}
|B \rangle = \sum_{n = 1}^qa_n |n \rangle
\end{equation}
and
\begin{equation}\label{eqe}
|E \rangle = \sum_{n = 1}^qa_n |N + 1 - n \rangle
\end{equation}
The energy splitting of the two lowest eigenvalues of $H_d(N)$ in a
finite chain is:
\begin{equation}
\Delta \lambda = 2 \langle B|H_d|E \rangle
\end{equation}
From (\ref{eqoffdiag}):
\begin{equation}
\langle i|H_d|j \rangle = H_d(|i - j|)
\end{equation}
Hence:
\begin{equation}\label{eqtaylor}
\langle B|H_d|E \rangle = \sum_{n,m = 1}^qa_n^*a_m \langle n|H_d |N
+ 1 - m \rangle = \sum_{n,m = 1}^qa_n^*a_m H_d(|N + 1 - m - n|)
\end{equation}
We adopt a `dummy' variable $X = |i - j|$, so that:
\begin{equation}
H_d(X) = \frac{C}{2a^3X^3}
\end{equation}
\begin{equation}
\frac{\partial H_d(X)}{\partial X} = -\frac{3C}{2a^3X^4}
\end{equation}
Using (\ref{eqtaylor}) and the fact that $L = a(N - 1)$, we can
expand $H_d(|N + 1 - m - n|)$ as a Taylor series to first order in
$\delta = m + n - 2$. Then:
\begin{equation}
H_d(|L - \delta|) = H_d(L) - \delta \left.\frac{\partial
H_d}{\partial X} \right|_{X = L} = \frac{C}{2L^3} + \frac{3Ca (m + n
- 2)}{2L^4}
\end{equation}
Hence:
\begin{eqnarray}
\langle B|H_d|E \rangle &=& \frac{C}{2}\left[ \frac{1}{L^3}\sum_{n,m
= 1}^q a_n^*a_m + \frac{a}{L^4}\sum_{n,m = 1}^q 3a_n^*a_m(m + n -
2)\right]\\ &=& \frac{C}{2} \left[\frac{Q}{L^3} +
\frac{aR}{L^4}\right]
\end{eqnarray}
with:
\begin{equation}\label{eqnq}
Q = \sum_{n,m = 1}^q a_n^*a_m
\end{equation}
\begin{equation}
R = \sum_{n,m = 1}^q 3a_n^*a_m (m + n - 2)
\end{equation}

\begin{figure}[h]
\renewcommand{\captionfont}{\footnotesize}
\renewcommand{\captionlabelfont}{}
\begin{center}
 \includegraphics[width=7cm]{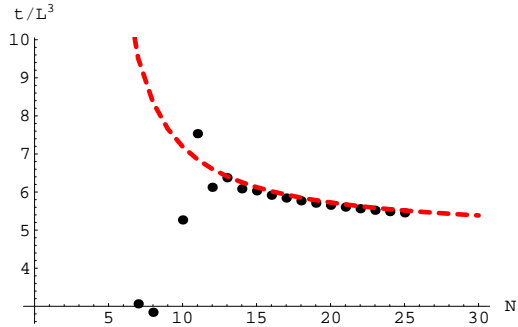}
 \end{center}
 \caption{(Colour Online) Comparison between the predictions of our q-spin model (red dashed points) and the
 data calculated from the treatment of the system in its entirety.
 We note the model becomes increasingly accurate at large
 \textit{N}.
 } \label{totfit}
\end{figure}

We find it is possible to model the asymptotic behaviour of the
system very accurately if we assign to the coefficients $a_n$ the
amplitudes of the ground state eigenvector of the $q \times q$
sub-matrix generated by truncating the full Hamiltonian $H_d(N)$ for
an arbitrarily large \textit{N} (fig.~\ref{totfit}). The values of
$a_n$ used to obtain the fit in fig.~\ref{totfit} come from the
ground state eigenvector of the $4 \times 4$ sub-matrix of
$H_d(14)$, and give $Q \approx 0.325$ and $R \approx -0.957$. These
quantities show only a very weak dependence on \textit{N}, so we
have assumed them to be constants. The fact that $Q < 1$ shows that
the transfer rate for chains of \emph{many} spins is always less
than that attained between two completely isolated spins; equation
(\ref{eqnq}) indicates this is a result of interference between
positive and negative components in the localized states $|B
\rangle$ and $|E \rangle$.

Therefore, only chains with \emph{few} spins can improve on the
performance of a simple dipole pair, as shown by the minimum in the
function $\tau(N)$ at $N = 4$ (fig.~\ref{normtimech}). This result,
together with fig.~\ref{mfch}, shows that, in a \emph{uniform}
chain, the best compromise between the quality and the speed of the
communication is obtained with 4 spins. This occurs because for
short chains the bound states at the two ends have a large overlap,
i.e. there exist terms in eqn.~(\ref{eqtaylor}) which simultaneously
have significant positive values of $a_n^*a_m$ and small values of
$|N + 1 - m - n|$. We have attempted to optimize the uniform 4-spin
chain still further, and find it is possible to improve its
performance slightly by modifying the positions of the inner spins
while maintaining mirror symmetry. For a chain of unit length, this
corresponds to taking $r_{1,2} = r_{3,4} \approx 0.314$ and $r_{2,3}
\approx 0.373$, which yield a value $\tau \approx 0.512$. However, a
comparison with $\tau \approx 0.568$ for a uniform chain ($r_{1,2} =
r_{2,3} = r_{3,4}\approx 0.333$) shows the improvement is minimal.

\subsubsection{Input and Output Optimization}
We now investigate the effects of altering the initial and final
states $|r \rangle$ and $|s \rangle$, while leaving the structure of
the chain intact. \begin{figure}[h]
\renewcommand{\captionfont}{\footnotesize}
\renewcommand{\captionlabelfont}{}
\begin{center}
 \includegraphics[width=8cm]{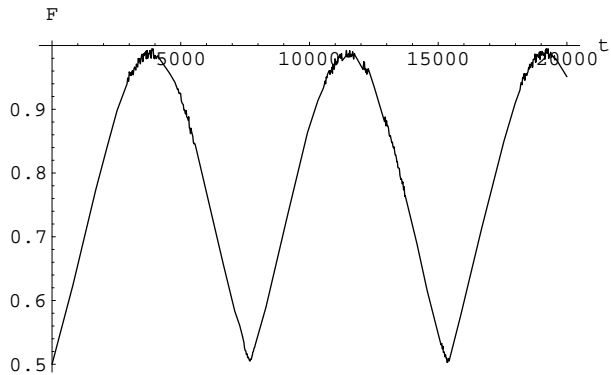}
 \end{center}
 \caption{The evolution in time (abscissa) of the fidelity of transmission of an input state
 of the form $C_1|1 \rangle + C_2|2 \rangle$ to an output state of the form
 $C_9|9 \rangle + C_{10}|10 \rangle$ in a uniform chain of 10 dipole-dipole coupled spins. Comparing with fig.~\ref{nine}, note that the curve is smoother
and the maximum fidelity has increased. Units are as specified in
Section \ref{The System}.
 } \label{superposnine}
\end{figure}\\If the starting and ending points are chosen at
random, the characteristic oscillation of $F_{r,s}^N(t)$ is lost,
unless either $|r \rangle = |2 \rangle$ and $|s \rangle = |N
\rangle$, or $|r \rangle = |1 \rangle$ and $|s \rangle = |N - 1
\rangle$. However, in both cases the signal is considerably noisier,
and the maximum fidelity is greatly reduced. This is a result of the
lesser efficiency of coupling to the bound states as one moves away
from the ends of the chain (cfr. fig.~\ref{spectrum}).

Conversely, it is possible to boost the maximum fidelity to unity
and smooth out all noise in the signal by encoding the states $|r
\rangle$ and $|s \rangle$ in two or more adjacent spins\footnote{The
possibility of improving state transfer by encoding a state in more
than a single spin is also discussed in
\cite{osborne04}.}(fig.~\ref{superposnine}). This is equivalent to
adopting $|r \rangle = |B \rangle$ and $|s \rangle = |E \rangle$,
where the $a_n$ are now obtained from the first and last \textit{n}
coefficients of the ground state eigenvector of $H_d(N)$, with the
additional condition that $\sum_n|a_n|^2 = 1$. This choice of input
and output states leaves the transfer time unaffected.

\section{Discussion and Conclusions}
We present a scheme for transferring quantum information through
infinite and finite chains of spins coupled via a pure magnetic
dipole interaction. This differs from much previous work in that the
dipole interaction is long-range, making for a system in which every
spin interacts with all other spins in the system, rather than with
nearest neighbours only. We find that, in general, the maximum
fidelity achievable by using a dipole-coupled system to transfer a
state between two maximally distant sites is greater or equal to
that which can be attained in a system exhibiting nearest-neighbour
interactions only.  The \emph{finite} chain, in particular, can be
engineered to give unit fidelity by simply adjusting the placement
of the spins and the input and output states. We have verified this
result only for $L = 2$ to $L = 23$, but believe it extends to
longer chains also.

The main weakness of both our systems is length of time taken to
complete the protocol, which increases polynomially in the size of
the system. However, we find that for a finite chain this obstacle
can be considerably lessened by simply modifying the relative
placement of the spins. Therefore, it does not necessarily preclude
the possibility of being able to transmit information over longer
distances on useful timescales. Furthermore, the protocol seems to
be reasonably robust against errors in spin placement; if we define
a ``failure rate" as the probability that the fidelity at time
$t(F_{max})$ will fall below the classical value, we find that, in a
uniform 4-spin chain of arbitrary length, a random error of 2 \% on
the placement of each spin yields a failure rate of approximately 5
\%. The corresponding uncertainty on $t(F_{max})$ is significantly
greater, but as $F_{1,N}^N(t)$ is a slowly varying function of time,
it is quite unlikely that the fidelity at $t(F_{max}) \pm \delta
t(F_{max})$ will have fallen significantly below the maximum.
Therefore, the simple and predictable behaviour of the fidelity in
time in a finite chain greatly increases the probability of carrying
out successful state transfer. The long range interactions also open
up the realistic possibility of measurements on individual spins,
and it would be interesting to investigate in the future whether
this can increase the speed of quantum state transfer as in
Ref.~\cite{vaucher05}.

\section*{Acknowledgements}
This work was supported by EPSRC through the Interdisciplinary
Research Collaboration in Quantum Information Processing.

\end{document}